\documentclass[seceq]{ptptex}

\usepackage{graphicx}
\usepackage{wrapft}
\usepackage{amsmath,amsbsy,amssymb,amscd,amsthm}

\title{%        %You can use \\ for explicit line-break.
Collective behavior of stock prices as a precursor to market crash%
}

\author{%       %Use \scshape for the family name.
Jun-ichi \textsc{Maskawa}%
}

\inst{%     %Affiliation, neglected when [addenda] or [errata].
Department of Economics, Seijo University\\6-1-20  Seijo, Setagaya-ku, Tokyo 157-8511, Japan
}

\abst{%         %This abstract is neglected when [addenda] or [errata].
We study precursors to the global market crash that occurred on all main stock exchanges throughout the world in October 2008 about three weeks after the bankruptcy of Lehman Brothers Holdings Inc. on 15 September. We examine the collective behavior of stock returns and analyze the market mode, which is a market-wide collective mode, with constituent issues of the FTSE 100 index listed on the London Stock Exchange. Before the market crash, a sharp rise in a measure of the collective behavior was observed. It was shown to be associated with news including the words "financial crisis." They did not impact stock prices severely alone, but they exacerbated the pessimistic mood that prevailed among stock market participants. Such news increased after the Lehman shock preceding the market crash. The variance increased along with the cumulative amount of news according to a power law.
}

\begin{document}

\maketitle

\section{Introduction}
What causes a drastic price change such as that occurring during a crash? According to the efficient market hypothesis, a prevailing paradigm of mainstream economics, stock prices change because of news that comes as a surprise to market participants. However, several previous works suggest that this picture of price changes is not empirically acceptable\cite{cutler,fair,joulin}. 

\begin{figure}[htbp]
\begin{center}
\includegraphics[bb=0 0 500 500,clip,width=12cm]{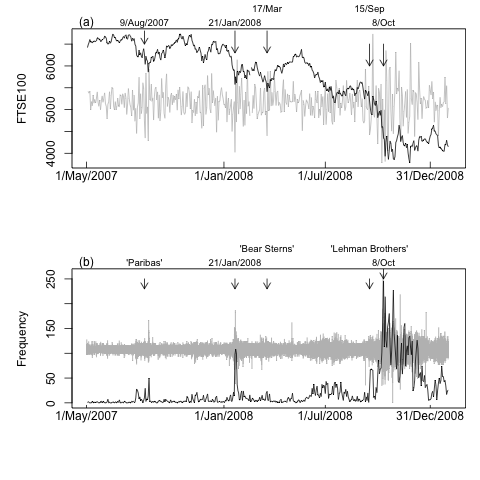}
\caption{Large price declines during May 2007 to Jan 2009. (a) Time series of the FTSE100 index (black line) and the daily log-return (gray line). (b) Evolution of the 1-min. log-return of Royal Dutch Shell plc (RDSA) (total:198490 min)(gray lines) and the daily frequency of large values with amplitudes greater than the twofold standard deviation during the period (black lines). Overnight price changes are excluded.}
\label{ftse100}
\end{center}
\end{figure}

From the Paribas' shock in August 2007 until the beginning of 2009, many stock markets throughout the world experienced considerable stock price declines because of ripple effects of the US sub-prime loan problem and the subsequent financial crisis. In Figure \ref{ftse100}(a), the time series of the FTSE100 index and the daily log-return are shown for the period from May 2007 through January 2009. Regarding some heavy falls such as that of the Paribas' shock on 9 Aug 2007, the buyout of Bear Stearns by JP Morgan on 17 March 2008, and the Lehman shock on 15 September 2008, they were probably caused by large exogenous shocks. However, larger falls such as the crashes of January and October 2008 are related to no specific news that might equal the magnitude of the subsequent drop-off. As an example, the evolution of the 1-min log-return of Royal Dutch Shell plc (RDSA) and the daily frequency of its large values of amplitudes larger than the twofold standard deviation during the period are also shown in Figure \ref{ftse100}(b)\footnote{Here, we exclude overnight price changes.}. It is apparent that the intermittent large price changes occurring during the daytime around 21 Jan. and 8 Oct. 2008 market crash without specific news are more frequent and persistent compared to the days showing steep declines that are apparently caused by important economic news. 

\begin{figure}[htbp]
\begin{center}
\includegraphics[bb=0 0 500 500,clip,width=10cm]{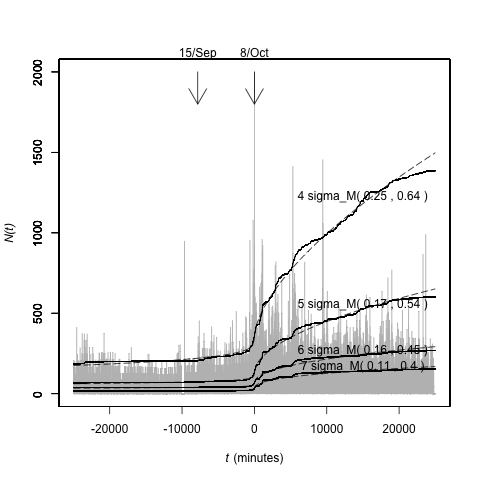}
\caption{Cumulative frequency of large 1-min. log-returns of RDSA. From top to bottom, we respectively show cumulative frequencies for the threshold $4\sigma_M, 5\sigma_M, 6\sigma_M, 7\sigma_M$ (black solid lines). Here $\sigma_M$ is the standard deviation during May 2007 -- Jan 2009. The dashed lines are the best power-law fits by Eq. (\ref{eq.omori}). The numbers in parentheses attached to each curve are $(\beta_b, \beta_a)$ for the corresponding threshold. The time-series of 1-min. log-return is also shown (gray line).}
\label{omori}
\end{center}
\end{figure}

Especially for the market crash of Oct. 2008, the daily frequency of large returns gradually grows from the Lehman shock on 15 Sep. until the maximum on 8 Oct. and decays slowly. Results demonstrate that the relaxation dynamics of a financial market immediately after the occurrence of a crash resemble an earthquake aftershock\cite{lillo}. The frequency of a large aftershock decays according to a power law. The power law of earthquake aftershocks was reported by Omori in 1984\cite{omori}. The market behavior before and after the scheduled macro-economic news announcements, such as U.S. Federal interest rate changes, has also been studied, revealing a similar law of foreshock\cite{petersen}. Figure \ref{omori} portrays a plot of the cumulative frequency $N(t)$ of large log-returns with amplitudes larger than a given threshold against the elapsed time $t$, which has an origin coordinate that is the time of the largest amplitude of the log-return during the daytime of 8 Oct. 

According to previous work by Petersen et al. \cite{petersen}, the cumulative frequency $N(t)$ is fitted by the following power function:

\begin{equation}
|N(t)-N(0)|\sim|t|^{\beta_i}~~(i=b,a),
\label{eq.omori}
\end{equation}
where the time regions $t<0$ ($i=b$) and $t>0$ ($i=a$) respectively correspond to a foreshock and aftershock. We have the inequality of $0<\beta_b<\beta_a<1$, which is the same result as that for the scheduled macro-economic news\cite{petersen}. Non-vanishing $\beta_b$ means that a foreshock occurs before the main shock.

Can we predict a singular point where the curvature of the power function change discontinuously at the main shock? In principle, it is impossible to determine the singular point because the estimated value $\beta_b$ would not be the value if the time of the main shock were not there. In this paper, we examine the collective behavior that affects the stock prices. 
We analyze the time-series of the average of normalized stock returns of stock index portfolio, which is approximately the first principal component of multivariate time-series of those returns and represents the market-wide collective behavior of stock prices\cite{gopik}. Designated as the "market mode" here, its rigorous definition is provided in the next section. 

As described in this paper, using the framework of the Multifractal Random Walk (MRW) model\cite{bacry}, we show the upwelling of collective behavior before the market crash. That is related to the news which includes the word "financial crisis", which has no severe impact to stock price alone but which contributes to worsening of the pessimistic mood among stock market participants. The amount of such news increased after the Lehman shock, which preceded the market crash. The variance increases along the cumulative amount of news according to a power law.

\section{Data}

We analyze the multivariate time-series of stock returns of 105 selected issues from the constituent of FTSE100 index listed on the London Stock Exchange for the period from May 2007 to January 2009 (445 days)\footnote{The constituents of FTSE100 index are updated frequently. We select those issues which had appeared on the list of the constituents at least once during the period and which had been listed on the London Stock Exchange throughout the period.}, which includes the period during which drastic price changes occurred because of the US sub-prime loan problem and the subsequent financial crisis. In Figure \ref{ftse100}(a), the time series of the FTSE100 index and the daily log-returns are shown for the period. 
We exclude the overnight price changes and specifically examine the intraday evolutions of returns\footnote{To average out the normalized returns of selected issues, we skip 60--110 min immediately after the market opening, varying from day to day until all the issues get opening prices.}. The total length of the time-series investigated here is $T=198,490$ min.
The return of the issue $i$ is the difference of the logarithms of log-price
of $t$ and $t-\delta t$:

\begin{equation}
\delta X_i(t)=\log(P_i(t))-\log(P_i(t-\delta t)).
\end{equation}
Here, we set  $\delta t=1$ and use the 1-min log-return. Next, we define the market mode $M(t)$ as the average of normalized returns:

\begin{equation}
\delta M(t)=\frac{1}{N}\sum^N_{i=1}\frac{\delta X_i(t)}{\sigma_i},
\end{equation}
where the number of issues $N=105$ and $\sigma_i$ is the standard deviation of log-return of i-th issue during the period. The market mode $\delta M(t)$ is approximately the first principal component of multivariate time-series of returns and represents the market-wide movement of stock prices\cite{gopik}. Finally, to remove the effect of the intraday U shape pattern of market activity from the time-series of the market mode, $\delta M(t)$ is divided by the standard deviation of the corresponding time of day, which is obtained from the whole of the time-series. We use the same notation $M(t)$ for the de-seasonalized market mode.

\section{Empirical results}

We use the framework of the Multifractal Random Walk (MRW) model\cite{bacry} to analyze the time-series of the de-seasonalized market mode $M(t)$.
We briefly introduce the MRW model in Appendix A, and show that the market mode is a multifractal random process and that it is well fitted by the model. The market mode is approximately the first principal component of multivariate time-series of $N$ returns and represents the market-wide collective behavior of stock prices\cite{gopik}. In the principal component analysis, the variance of principal components expresses the contribution to all the variation of multivariate time-series. From equations (\ref{eq.cov}) and (\ref{eq.rho}), the variance of $\omega_{\delta t}$ the logarithm of the volatility is expressed by $\lambda^2\log(L/\delta t)$. We regard the variance $Var(\omega_{\delta t})$ as the quantity representing the intensity of the market-wide collective behavior, i.e. herding of market participants.

\begin{figure}[htbp]
\begin{center}
\includegraphics[bb=0 0 500 500,clip,width=8cm]{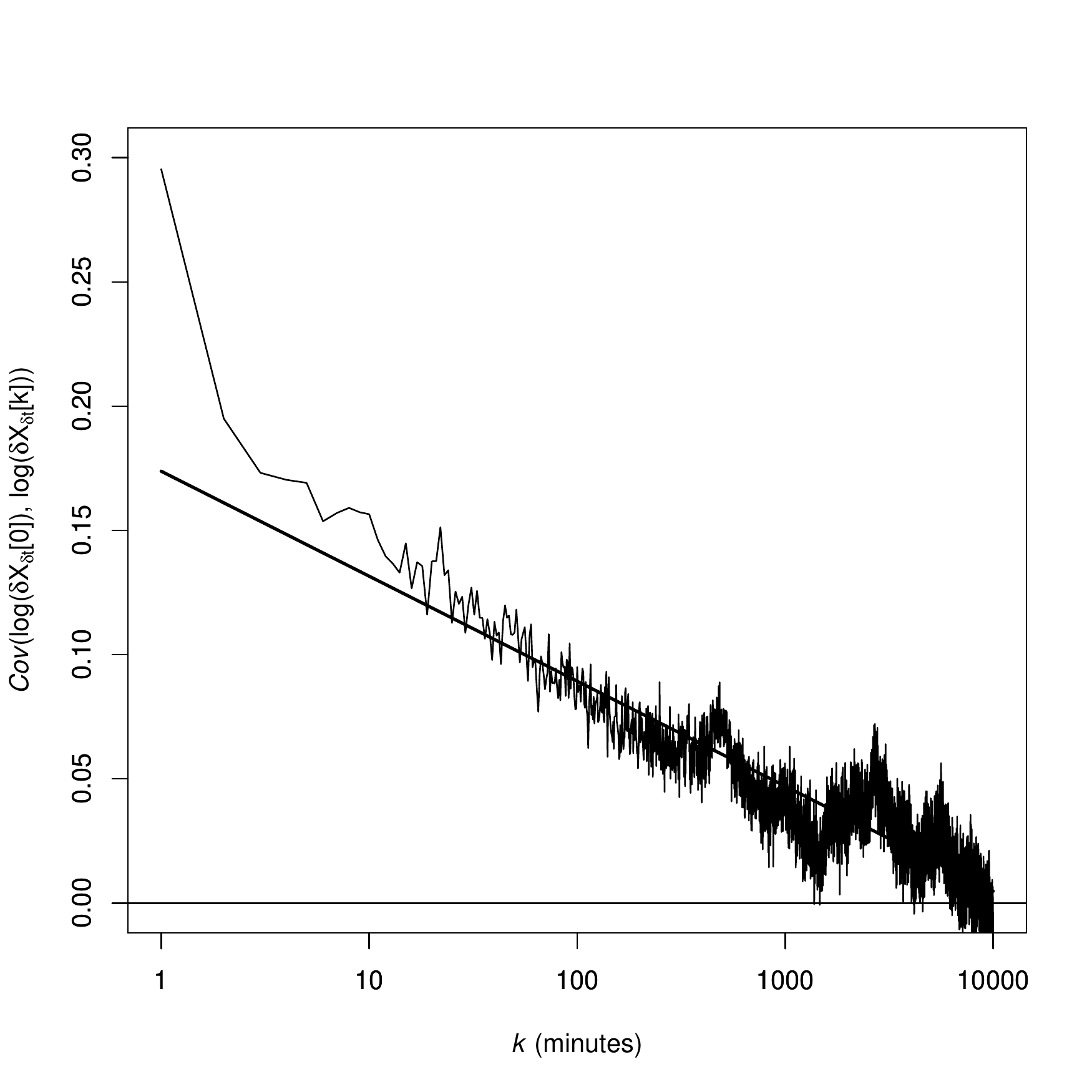}
\caption{Example of covariance function for a time window. The semi-logarithmic plot of $Cov(\log(\delta X_{\delta t}[0]),\log(\delta X_{\delta t}[k]))$ against lag k is shown. The straight line is the prediction of the MRW model (\ref{eq.covl}). The estimated values are $\lambda^2=0.018$ and $L=12975.43$.}
\label{acf}
\end{center}
\end{figure}

In the MRW model, $\omega_{\delta t}$ is a stationary process. Therefore, the parameters $\lambda$ and $L$ are constant against time. However, the intensity of the market-wide collective behavior is thought to undergo a change depending on the market phase. We therefore evaluate the temporal behavior of the variance $Var(\omega_{\delta t})$ in a sliding time window $[t-\Delta T,t ]$ with given width of $\Delta T$\cite{kiyono}. 
To obtain the variance $Var(\omega_{\delta t})$ in each window, we must estimate parameters $\lambda$ and $L$. From analytical results of MRW model, the covariance of the logarithm of absolute return follows the equation\cite{bacry},

\begin{equation}
Cov(\log(\delta X_{\delta t}[i]),\log(\delta X_{\delta t}[i+k])) = 
\begin{cases}
-\lambda^2\log(|k|/L) & for~\delta t\ll |k|\le L \\
0 & L < |k|.
\end{cases}
\label{eq.covl}
\end{equation}
In Figure \ref{acf}, an example of the covariance function for a time window is shown. The covariance function is fitted by the prediction of the MRW model (\ref{eq.covl}). Parameters $\lambda$ and $L$ are estimated using this equation. We set width $\Delta T$ a fifth of the whole period, namely, $\Delta T=39698$ min, which is in the range $L < \Delta T < T$.
 
\begin{figure}[htbp]
\begin{center}
\includegraphics[bb=0 0 500 500,clip,width=10cm]{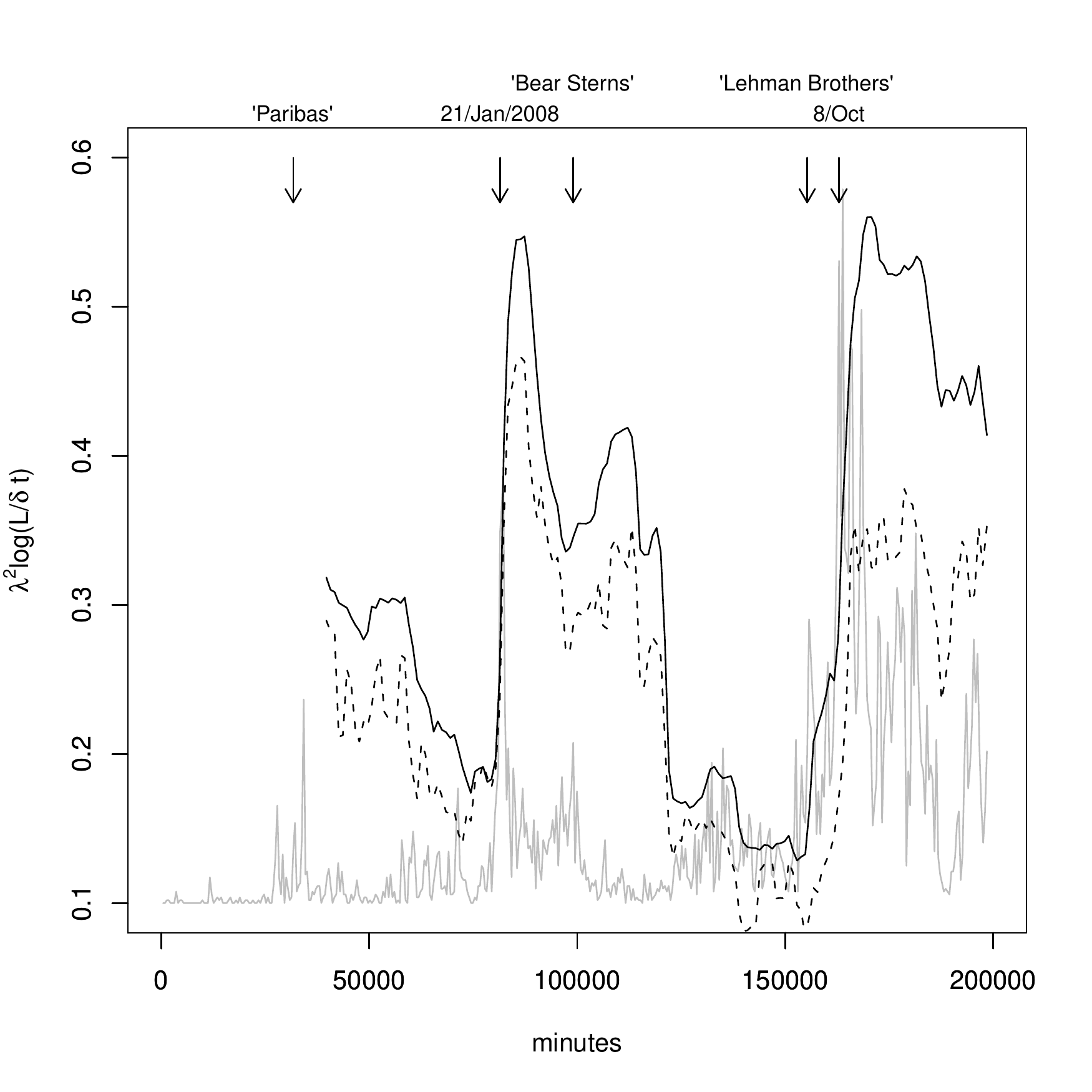}
\caption{Temporal evolution of the estimated variance $\lambda^2\log(L/\delta t)$. Two results are shown. One is the result with no de-trending prescription and $\delta t=1$ min (black solid line). The other is the result with the thorough prescription described in footnote and $\delta t=8$ min (dashed line). The daily frequency of large $\delta M(t)$ values with amplitudes larger than the twofold standard deviation during the period (black lines) is also shown (gray line).}
\label{lambda}
\end{center}
\end{figure}

In Figure \ref{lambda}, we show the temporal evolution of the variance $\lambda^2\log(L/\delta t)$.
In general, for a non-stationary process, de-trending of the time-series must be prescribed. However, the de-trending method of is not determined uniquely. Therefore, we compare the result without de-trending with that using a thorough de-trending prescription\footnote{Here we remove the local trend every 8 min as the thorough de-trending prescription.}.
 In both cases, the sharp rises of the variance $\lambda^2\log(L/\delta t)$, which mean the upwelling of the collective behavior of stock prices, are observed before the market crashes of Jan. and Oct. 2008. The periods of high volatility lasted several months after the crashes. This phenomenon is not observed for steep declines of Mar. and Sep. 2008, which were caused by the buyouts of Bear Stearns by J.P. Morgan on 17 March and the Lehman shock on 15 September, respectively.

\begin{figure}[htbp]
\begin{center}
\includegraphics[bb=0 0 500 500,clip,width=10cm]{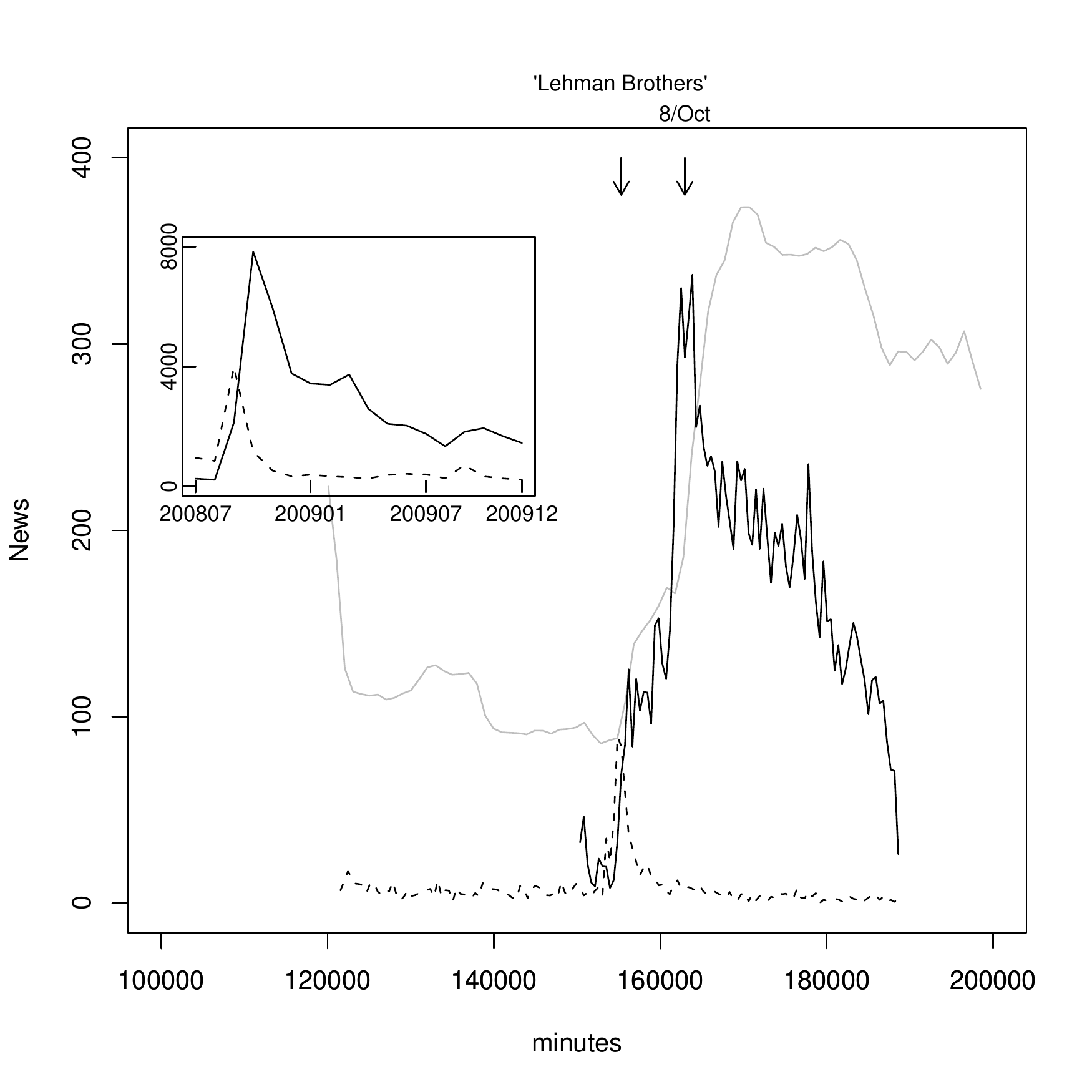}
\caption{Temporal evolution of the daily amounts of news which include the words "financial crisis" of the period from 1 Sep. to 31 Dec. 2008 (black solid line), the news with the words "Lehman Brothers" of the period from 1 Jul. to 31 Dec. 2008 (dashed line)\cite{factiva}. The news sources are confined to Reuters news. The amounts of news strongly depends on the day of the week. To remove that effect, we divided the daily amount by the average value of the corresponding day of the week and multiply it by the average during the period. The temporal evolution of the variance $\lambda^2\log(L/\delta t)$ with no de-trending prescription is also shown (gray line). Inset: Actual monthly count of both news of the period from Sep. 2008 to Dec. 2009. }
\label{news}
\end{center}
\end{figure}

As the final empirical result, we demonstrate that the sharp rise of the variance before the crash of Oct. 2008 is associated with the news which include the words "financial crisis." They have no severe impact on stock return alone but contribute to the pessimistic mood among stock market participants. The amount of such news increased after the Lehman shock preceding the market crash (Figure \ref{news}). The dynamics of the amount of news with specific words before and after the crash is well described by the precursory and the relaxation dynamics of a social system, which has been studied in \cite{crane} using time-series of daily views of video on YouTube. News including the words "financial crisis" increased toward the peak, taking about three weeks and showing slow relaxation lasting over the end of 2009, which is characteristic of an "endogenous" burst. In contrast, the news with the words "Lehman Brothers" show a sudden peak and rapid relaxation as an typical profile of "exogenous" burst\cite{crane}. 

\begin{figure}[htbp]
\begin{center}
\includegraphics[bb=0 0 500 500,clip,width=8cm]{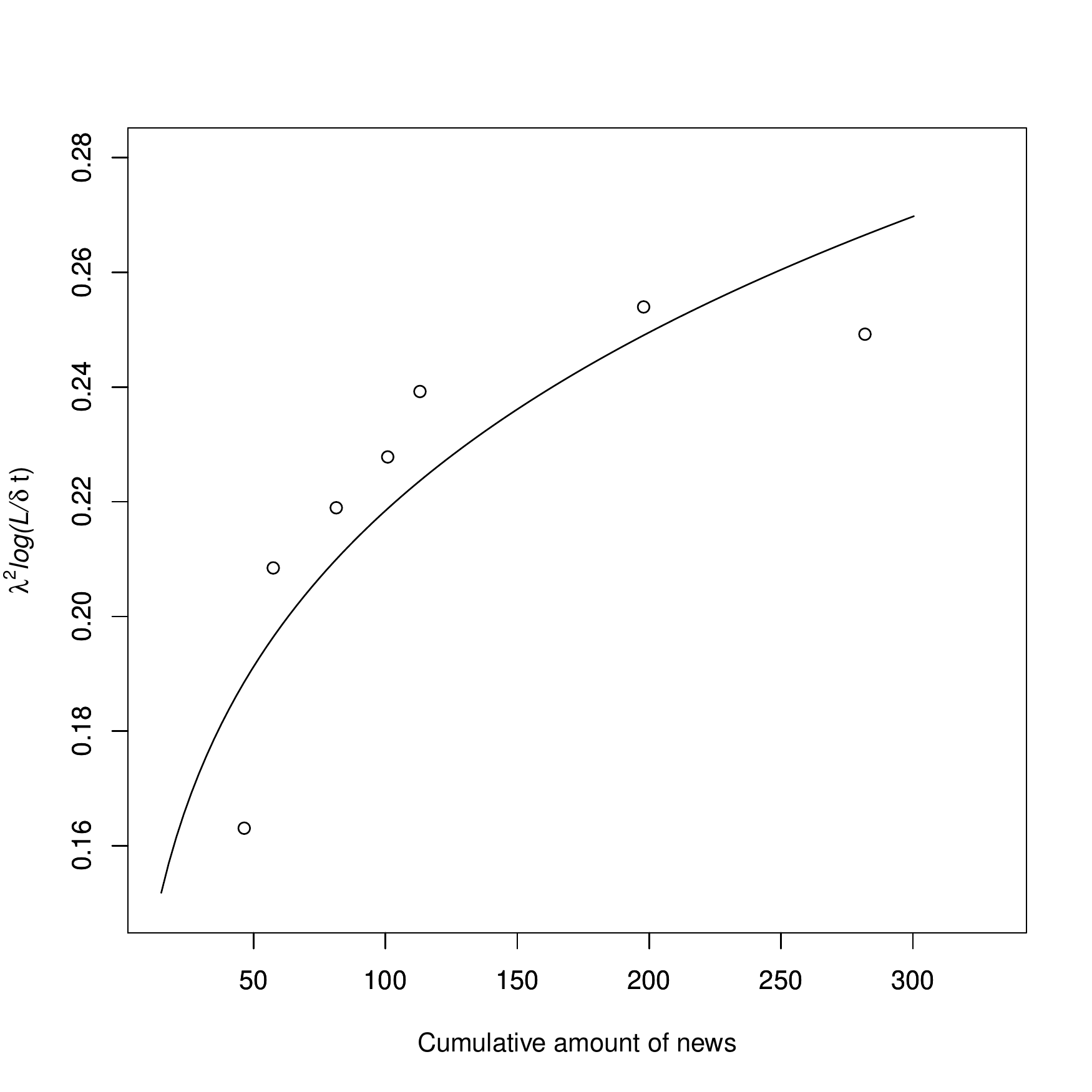}
\caption{Growth of the variance $Var(\omega_{\delta t})=\lambda^2\log(L/\delta t)$ against the cumulative amount of news $N_n$for the period from 16 Sep. to 8 Oct (circle). The solid line represents the best fit by eq. (\ref{eq.power_law}). The estimated exponent $\alpha_n$ is $0.19\pm0.04$ }.
\label{lambda_vs_news}
\end{center}
\end{figure}

Figure \ref{lambda_vs_news} shows the increase of the variance $Var(\omega_{\delta t})=\lambda^2\log(L/\delta t)$ against the cumulative amount of news $N_n$ for the period from 16 Sep. to 8 Oct., which is expressed by the power-law 

\begin{equation}
Var(\omega_{\delta t})(t) \sim N_n(t)^{\alpha_n},
\label{eq.power_law}
\end{equation}
where the estimated exponent $\alpha_n=0.19\pm0.04$ in this case.

\section{Conclusions}

As described in this paper, we have shown through an empirical study of the stock returns of the constituent issues of FTSE 100 index listed on London Stock Exchange for the period from May 2007 through Jan. 2009 that precursors of the market crash of Jan. and Oct. 2008 exist. A sharp rise preceding the crashes in the collective behavior of stock prices was measured by the variance of the logarithm of return, which is a parameter of the MRW model. This phenomenon signifies the upwelling of the market-wide collective behavior before the crash, which might reflect a herding of market participants. The crash of Oct. 2008 was synchronized with the increase of news including the words "financial crisis." The variance increases along with the cumulative amount of news according to a power law.

It has remained unknown how to predict the time and the magnitude of steep decline such as that of a market crash. However, it might be accomplished through the accumulation of understanding about complex systems.

\section*{Acknowledgements}
\noindent
The author thanks Koji Kuroda and Joshin Murai for helpful discussions, and thank Sayaka Tamura for her downloading of news data including the words "Lehman Brothers". 
He also thanks the Yukawa Institute for Theoretical Physics at Kyoto University.
Discussions during the YITP Workshop YITP-W-11-04 on "Econophysics 2011 -- The Hitchhiker's Guide to the Economy" were useful to complete this work. This research was partially supported by a Grant-in-Aid for Scientific Research (C) No. 20510153.

\appendix
\section{Multifractal Random Walk (MRW) model}

Multifractal Random Walk (MRW) $X(t)$ is a continuous process defined by the limit of the discrete random process $X_{\delta t}$:

\begin{equation}
X(t)=\lim_{\delta t \to 0,t=K_{\delta t}\delta t}X_{\delta t}(K_{\delta t}\delta t).
\end{equation}
The discrete process $X_{\delta t}(K_{\delta t}\delta t)$ is a stochastic volatility process that can be decomposed into subprocesses $\delta X_{\delta t}$:

\begin{equation}
X_{\delta t}(K_{\delta t}\delta t)=\sum_{i=1}^{K_{\delta t}}\delta X_{\delta t}[i].
\end{equation}
The subprocess $\delta X_{\delta t}(i)$ is described as

\begin{equation}
\delta X_{\delta t}[i]=\epsilon_{\delta t}[i]\exp({\omega_{\delta t}[i]}),
\end{equation}
where $\epsilon_{\delta t}$ is a stationary Gaussian white noise with variance $\sigma^2 \delta t$ and $\exp({\omega_{\delta t}[i]})$ is the stochastic volatility.

Bacry et al. show that the stochastic volatility process $X_{\delta t}(i)$ is a multifractal process\cite{bacry}, as

\begin{equation}
M(q,\delta t)=E(|\delta X_{\delta t}|^q)\sim \delta t^{\zeta_q}
\label{eq.mql}
\end{equation}
if $\omega_{\delta t}[i]$ is a stationary Gaussian process such that $E(\omega_{\delta t})=-Var(\omega_{\delta t})$ and

\begin{equation}
Cov(\omega_{\delta t}[i],\omega_{\delta t}[j])=\lambda^2 \log \rho _{\delta t}[|i-j|],
\label{eq.cov}
\end{equation}
where

\begin{equation}
\rho_{\delta t}[k] = \begin{cases}
\frac{L}{(|k|+1)\delta t} & for~|k|\le L/\delta t -1\\
1 & otherwise.
\end{cases}
\label{eq.rho}
\end{equation}
Spectrum $\zeta_q$ is given by the formula

\begin{equation}
\zeta_q=(q-q(q-2)\lambda^2)/2.
\label{eq.zeta}
\end{equation}
To apply a MRW model to a time-series, three parameters $\sigma, \lambda$, and $L$ must be estimated.

\begin{figure}[htbp]
\begin{center}
\includegraphics[bb=0 0 500 500,clip,width=10cm]{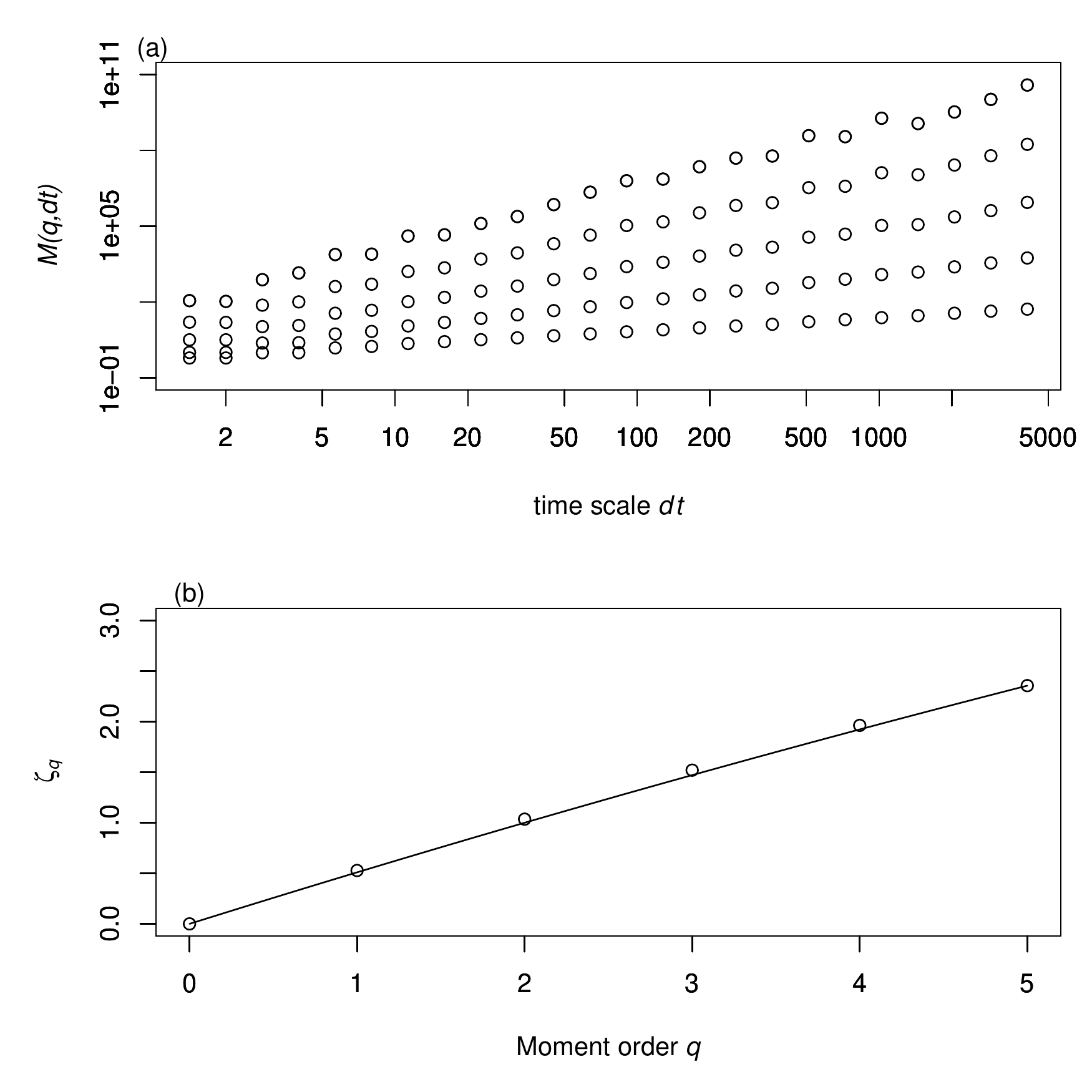}
\caption{Multifractal properties of the time-series of the market mode. (a) Double logarithmic plot of the expectation values of q moments against the time scale $\delta t$ is shown. From top to bottom, the moment order $q=1, 2, 3, 4, 5$. Time scale $\delta t$ varies from 1 min. to 4096 min. (b) Spectrum $\zeta_q$ (circle) and the best fit by the theoretical prediction (\ref{eq.zeta}) (solid line) are shown.}
\label{mql}
\end{center}
\end{figure}

Here, we check the applicability of MRW model to the market mode $M(t)$ using the formulae of (\ref{eq.mql}) and (\ref{eq.zeta}).
Figure \ref{mql} shows the double logarithmic plot of the expectation values of $q$ moments against the time scale $\delta t$, and the spectrum $\zeta_q$, which is estimated by eq. (\ref{eq.mql}). Those results mean that the MRW model is well applicable to the time series of the de-seasonalized market mode.

\end{document}